\documentclass[utf8]{frontiersSCNS} 

\usepackage{url,hyperref,lineno,microtype,subcaption}
\usepackage[onehalfspacing]{setspace}
\usepackage{aas_macros}

\def\keyFont{\fontsize{8}{11}\helveticabold }
\def\firstAuthorLast{Gilbert {et~al.}}
\def\Authors{Emily A. Gilbert\,$^{1,2,3,4*}$, Thomas Barclay\,$^{2,4}$, Ethan Kruse\,$^{4}$, Elisa V. Quintana\,$^{4}$ and Lucianne M. Walkowicz\,$^{3}$}

\def\Address{$^{1}$, Department of Astronomy and Astrophysics, University of
Chicago, Chicago, IL, USA\\
$^{2}$Center for Space Science and Technology, University of Maryland, Baltimore County, Baltimore, MD, USA\\
$^{3}$The Adler Planetarium, Chicago, IL, USA\\
$^{4}$Exoplanets and Stellar Astrophysics Laboratory, NASA Goddard Space Flight Center, Greenbelt, MD, USA
}

\def\corrAuthor{Emily A. Gilbert, Department of Astronomy and Astrophysics, University of
Chicago, 5640 S. Ellis Ave, Chicago, IL 60637, USA}

\def\corrEmail{emilygilbert@uchicago.edu}

\begin{document}
\onecolumn
\firstpage{1}

\title[No Transits of Proxima Centauri Planets]{No Transits of Proxima Centauri Planets in High-Cadence TESS Data} 

\author[\firstAuthorLast ]{\Authors} 
\address{} 
\correspondance{} 

\extraAuth{}

\maketitle

\begin{abstract}
Proxima Centauri is our nearest stellar neighbor and one of the most well-studied stars in the sky. In 2016, a planetary companion was detected through radial velocity measurements. Proxima Centauri b has a minimum mass of 1.3 Earth masses and orbits with a period of 11.2 days at 0.05 AU from its stellar host, and resides within the star's Habitable Zone. While recent work has shown that Proxima Centauri b likely does not transit, given the value of potential atmospheric observations via transmission spectroscopy of the closest possible Habitable Zone planet, we reevaluate the possibility that Proxima Centauri b is a transiting exoplanet using data from the Transiting Exoplanet Survey Satellite (TESS). We use three sectors (Sectors 11, 12, and 38 at 2-minute cadence) of observations from TESS to search for planets. Proxima Centauri is an extremely active M5.5 star, emitting frequent white-light flares; we employ a novel method that includes modeling the stellar activity in our planet search algorithm. We do not detect any planet signals. We injected synthetic transiting planets into the TESS and use this analysis to show that Proxima Centauri b cannot be a transiting exoplanet with a radius larger than 0.4 R$_\oplus$. Moreover, we show that it is unlikely that any Habitable Zone planets larger than Mars transit Proxima Centauri.

\tiny
 \keyFont{ \section{Keywords:} Stellar activity --- 
M dwarf stars --- Stellar flares --- Exoplanet detection methods --- Algorithms} 

\end{abstract}
\section{Introduction}\label{sec:intro}

M dwarf stars' small sizes make them particularly advantageous when it comes to exoplanet detection via both radial velocity and transit methods. However, these low-mass stars are frequently highly magnetically active for long portions of their lifetimes, which can confound detection and further characterization of planets orbiting such stars. White light flares may induce radial velocity shifts, increasing noise in RV measurements and hindering planet detections \citep{Saar1997,Korhonen2015,Andersen2015}. Furthermore, white light flares may also contaminate or even completely mask planetary transits in light curves, once again complicating planet detection \citep{Oshagh2013,Gilbert2021}.

Proxima Centauri (Proxima, Prox Cen) is the nearest star to our own Solar System and a well known flare star \citep{Thackeray1950,Shapley1951}. Proxima Centauri is an M5.5 dwarf star with an estimated age of $\sim$5 Gyrs \citep{Thevenin2002}. Proxima Centauri is extremely active, producing near-continuous flares at a variety of wavelengths \citep[e.g.][]{macgregor2021}. 
Owing to its proximity, Proxima Centauri is one of the most well-studied stars in the sky, and as such, has been a prominent target for exoplanet searches for many years \citep[e.g.][]{Holmberg1938,Benedict1999,kurster1999,Endl2008,Lurie2014}. 

Despite the challenges posed by observing an active star, \citet{discovery} used radial velocity measurements to uncover Proxima Centauri b, a minimum mass 1.3 M$_{\oplus}$ planet orbiting Proxima Centauri with a period of 11.2 days. With an orbital semi-major axis of 0.05 AU, Proxima Centauri b has an estimated T$_{eq}$ of T = 234 K \citep{discovery} and resides in the star's Habitable Zone \citep{kopparapu2013}. A tentative detection of an outer companion, Proxima Centauri c, was announced later in 2020 \citep{Damasso2020}, once again detected through radial velocity. Proxima Centauri c has a minimum mass of 5.8$\pm$1.9M$_\oplus$ and an orbital separation of 1.48 AU with a period of 5.2 years. Both of these planets were found via the radial velocity method, but from the geometry of the system, it is possible to determine the probability of transits of either planet occurring. Given its orbital separation from its host star, the geometric probability that planet b transits is around 1.5\% \citep{discovery}, and at 1.5 AU, the probability of planet c transiting is negligible.

While neither planet is known to transit, if either planet (or another yet undiscovered planet) were found to transit, the prospects for follow-up would be significant. A precise radius would allow for accurate determination of the planets' compositions, and a transit would enable transmission spectroscopy observations. Furthermore, given the location of Proxima Centauri b within the Habitable Zone of its host star, and its short orbital period, which would provide ample transit opportunities, it would be an ideal target to search for the presence of biosignatures. Missions such as the upcoming James Webb Space Telescope (JWST) would be able to confirm the presence of an atmosphere on Proxima Centauri b if it does in fact have one \citep{kreidberg2016} through thermal phase curve analysis.

Given the significance of a nearby, Habitable Zone planet, since the discovery of Proxima Centauri b, a number of teams have studied Proxima Centauri to definitively rule out Proxima b as a transiting exoplanet. \citep{discovery} found no evidence of transits in all available photometric light curves and noted a 1.5\% geometric probability of transit.

\citet{kipping2017} observed Proxima Centauri for 43.5 days from 2014 - 2015 with \textit{MOST} space telescope. While they were unable to recover planet b in the photometry, they did see two candidate transit signals with the expected depth of a transiting planet the size of Proxima Centauri b. Their efforts were hindered by the frequent flaring of the host star, and they determined their false-negative detection rate to be 20-40\%.

\citet{li2017} used a robotic 30 cm telescope at Las Campanas Observatory to observe Proxima for 23 nights in 2016. They found inconclusive evidence for transits of planet b, and they also found an additional transit candidate that was inconsistent with the transit window of Proxima b, perhaps indicating an additional planet.

\citet{liu2018} used the Bright Star Survey Telescope at the Zhongshan Station in Antarctica to observe Proxima Centauri, conducting a 10-day long photometric monitoring campaign in 2016. This work also detected a transit-like signal consistent with the RV orbital parameters of planet b, but the researchers were unable to definitively confirm a transit of Proxima b.

\citet{Blank2018} conducted a multi-year search for transits of Proxima Centauri b spanning 2006--2017. This search utilized of 329 observations from telescopes across the world and they were unable to verify the previous claims of tentative transit detections corresponding to Proxima Centauri b, although they did confirm the presence of significant photometric variability of the host star due to flares. \citet{Feliz2019} built off of this work and conducted a transit search for periods ranging from 1-30 days using 262 high-quality light curves from the \citet{Blank2018} analysis. They conducted a series of planet injection tests and found that their methods were sufficiently sensitive to have detected transits of 5 millimagnitude depth and saw no evidence of periodic transit-like events over the 1-30 day period range. However, due to incomplete phase coverage of the light curves and a lack of sensitivity to transits shallower than 4 millimagnitudes, they could not rule out a detection of transits of Proxima Centauri b.

\citet{Jenkins2019} used 48 hours of \textit{Spitzer} data to search for transits of Proxima b at 4.5 microns, going further to the infrared in an effort to mitigate the effects of stellar activity. With this data, they were able to rule out planetary transits at the 200 ppm level, putting a 3$\sigma$ upper limit on the radius of Proxima Centauri b at 0.4 R$_\oplus$.

Many of the above searches indicate that the frequent flaring activity hindered their transit searches and may be masking transits of planet b, or other planets in the system. While the tentative detections may be the results of transits, they may also be the result of atmospheric variability during ground based observations, treatment of flares, or stellar variability. 

The Transiting Exoplanet Survey Satellite (TESS) mission \citep{ricker14} has revolutionized our ability to study white light flares in great detail. TESS aims to observe the entire sky at optical/near-IR wavelengths in order to search for nearby, transiting exoplanets in high-cadence photometry. For 200,000 targets, spanning nearly the entire sky, TESS has collected high-precision photometry at 2-minute cadence, allowing us an unprecedented window into the Proxima Centauri system.

In this work, we present a new analysis using TESS data to search for transits of Proxima Centauri b. In Section \ref{sec:data}, we describe the TESS observations used in this analysis. Next, we describe our data reduction, transit search, and injection and recovery test in Section \ref{sec:methods}. We present a discussion of these results in Section \ref{sec:discussion}, and conclude with a summary in Section \ref{sec:conclusions}.

\section{Data and Observations} \label{sec:data}

TESS observed Proxima Centauri in 2019 from April 23 -- June 18 during observing Sectors 11 and 12, and then again from April 29 -- May 26, 2021 in observing Sector 38~\footnote{Proxima Centauri was observed in Sectors 11, 12 and 38 through the inclusion of the star in 16 Guest Investigator programs: G011129 (PI: Jao, Wei-Chun), G011153 (PI: Pineda, J. S.), G011180 (PI: Dressing, C.), G011183 (PI: Kane, S.), G011185 (PI: Davenport, J.), G011231 (PI: Winters, J.), G011264 (PI: Davenport, J.), G011266 (PI: Schlieder, J.), G03273 (PI: Vega, L.), G03225 (PI: Pineda, J. S.), G03205 (PI: Monsue, T.), G03174 (PI: Howard, W.), G03250 (PI: Winters, J.), G03226 (PI: Silverstein, Mi.), G03202 (PI: Paudel, R.), G03227 (PI: Davenport, J.).}. 

The Sector 11 observations totalled 26.04 days between orbits 29 and 30, from 2019 April 23 (1596.77203 TESS Julian Date, TJD)\footnote{TESS timestamps are Barycentric Julian Date - 2457000.} to 2019 May 20 (1623.89147 TJD) with 1.08 days of downtime in observations for data downlink using Camera 2, CCD 2. 

Sector 12 observations of Proxima occurred during TESS orbits 31 and 32, on Camera 2, CCD 1, from  2019 May 21 (1624.94979 TJD) to 2019 June 18 (1652.89144 TJD) for a total of 26.90 days of science data with a 1.04 day gap for data downlink. 

TESS Sector 38 observations of the target were conducted from 2021 April 29 (2333.84945 TJD) to  2021 May 26 (2360.55083 TJD) during orbits 83 and 84 for 25.74 days of science with a 0.96 day gap between the orbits to download data. These observations of Proxima were taken on Camera 2, CCD 2.

The 2019 data were collected at 2-minute cadence and the 2021 data at both 2-minute and in the new 20-second cadence mode. These data were processed by the TESS Science Processing Operations Center (SPOC) pipeline \citep{jenkinsSPOC2016}, and archived at the the Mikulski Archive for Space Telescopes (MAST). We retrieved the light curves used in this analysis from MAST using \texttt{lightkurve} \citep{lightkurve}. Proxima Centauri is bright in the TESS bandpass at T$_{\text{mag}}$=7.6, providing an ideal target to search for planets.

\section{Methods}
\label{sec:methods}

Proxima Centauri is an active star, emitting frequent white light flares. We see 2-3+ large flares every day in the 2-minute cadence TESS light curve \citep{Vida2019}, and even with TESS 2-minute cadence, optical photometry, it can be hard to fully resolve flare morphology in order to search for transits. \citet{Davenport2016} even suggest that the visible-light light curve of Proxima Centauri may be so dominated by flares that the time series can be thought of as primarily a superposition of many flares. This level of activity can complicate the search for transiting exoplanets because of the additional noise in the data. A typical method of dealing with large flares is to identify them and remove them either by using flare detection algorithms or simple sigma-clipping. Here we take a different approach, we identify the flares using a custom algorithm, model the flares using a template, subtract these flares from the data, and then perform the transit search. We then inject transits into the light curve to test our sensitivity to transiting planets. 

\subsection{Flare Detection and Modeling}\label{sec:flares}

\begin{figure}
    \centering
    \includegraphics[width=
    \textwidth]{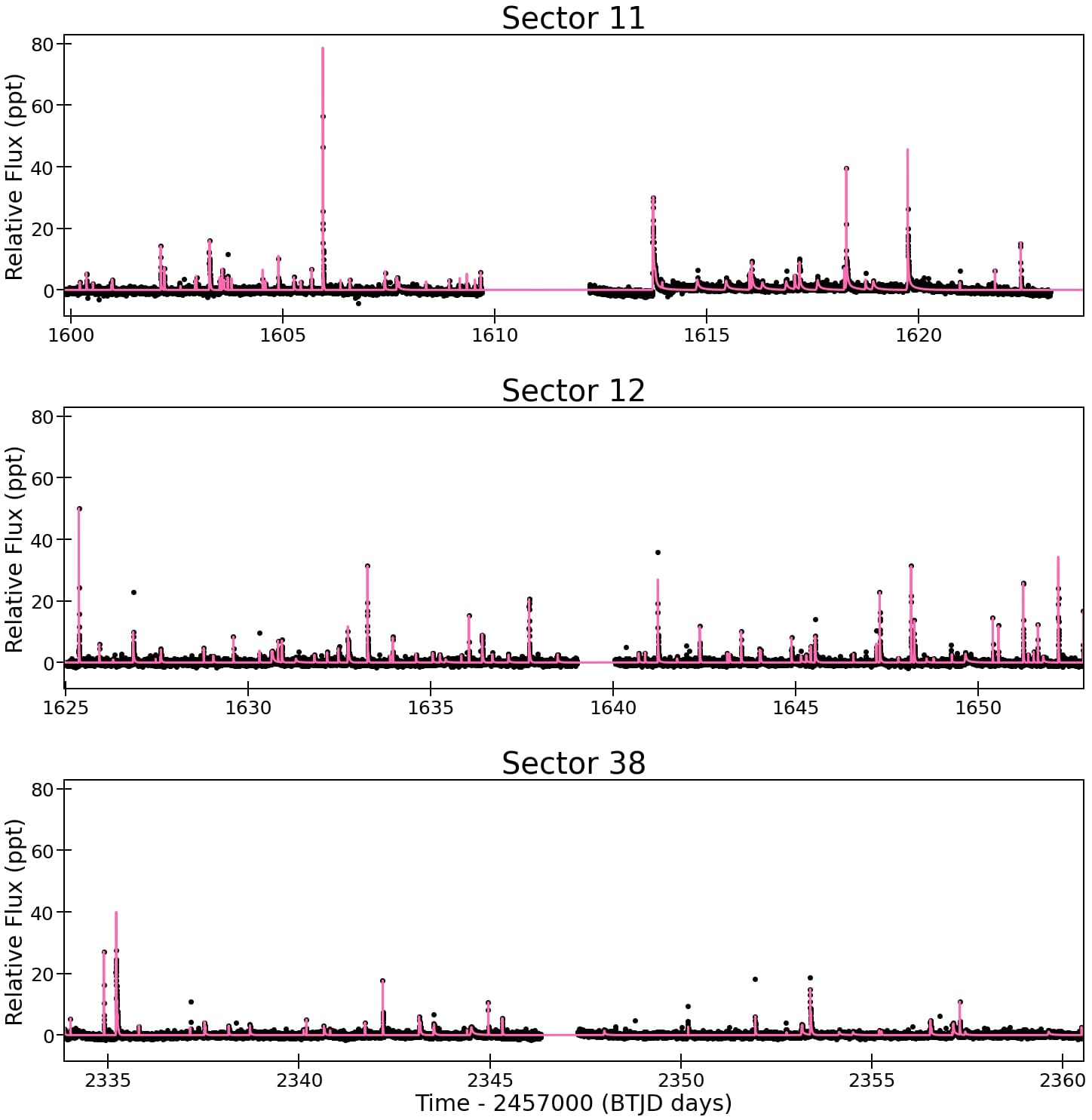}
    \caption{Here we show the raw TESS light curves for each sector (black) with our flare model overlaid (pink). By carefully modeling the flares and subtracting them off, we are able to conduct better transit searches in the data.}
    \label{fig:lightcurves}
\end{figure}

We detected flares using an updated version of \texttt{BayesFlare} \citep{pitkin14} modified to process TESS data. \texttt{BayesFlare} is an automated flare detection package that identifies flares in photometric light curves using Bayesian inference. For this look at Proxima Centauri, we used the 2-minute cadence TESS data for Sectors 11, 12, and 38. Although 20-second cadence data exist for Sector 38, we elected to only use the 2-minute cadence light curves for continuity and because 20-second data does not provide additional signal-to-noise in the transit search. 

After identifying flares automatically with \texttt{BayesFlare}, we opted to correct poor flare fits evidenced by large residuals. We did this by manually adjusting initial guesses of flare parameters (primarily duration that were too long) and adding flares that may not have been detected through the flare detection routine. This included adding peaks complex flares to ensure they were fit properly (62 total flares were added). In total we included 201 flares in the model (67 in Sector 11, 78 in Sector 12, and 56 in Sector 38), and for each flare (or component of a complex flare) we have estimates of the flare peak-time, full-width at half-maximum (FWHM), and amplitude. 

We then used \texttt{xoflares} \citep{xoflares} to measure the properties of the flares. Typically, \texttt{xoflares} would use the \texttt{PyMC3} framework to performing probabilistic sampling of flare properties. However, here we are less concerned with understanding the precise properties of the flares, but rather, just need a ``good enough" model that can be calculated rapidly -- our goal is not to be statistically principled in the flare modeling. Therefore, we calculated a \textit{maximum a priori} (MAP) model through an optimization of the flare parameters and use that in our analysis. The model was parameterized in terms of the flare peak time, the inverse of the FWHM and the amplitude, and a mean level. The priors on the MAP fit are Gaussian for all parameters with a mean from the initial flare search and standard deviation of 0.003 days for the peak time, and 50\% of the mean estimate for the inverse FWHM and amplitude. The flare amplitudes and FWHMs are constrained to be positive. We use a Student-t likelihood of the data rather than a Gaussian, because it is more forgiving of outliers.

After calculating a MAP model we subtracted it from the observed data and then sigma-clipped large outliers. These outliers were usually the result of a sub-optimal fit to the flare amplitude (a few are visible in Figure~\ref{fig:lightcurves}). Data more than to +5/-18-$\sigma$ from the median of the MAP model subtracted data are clipped, where the asymmetry comes from not wanting to clip any transits. An 18-$\sigma$ outlier corresponds to a 1.5 R$_\oplus$ planet, the largest radius we search for in our injection and recovery test. Then we calculated a second order Savitzy-Golay filter with a window-length of 1.1 days, and subtracted it from the original data (flares include) to capture and remove long-term trends. We then went through the flare fitting and sigma-clipping steps a second time and were left with the `quiescent' light curve that we used to search for planets.

\begin{figure}
    \centering
    \includegraphics[width=
    \textwidth]{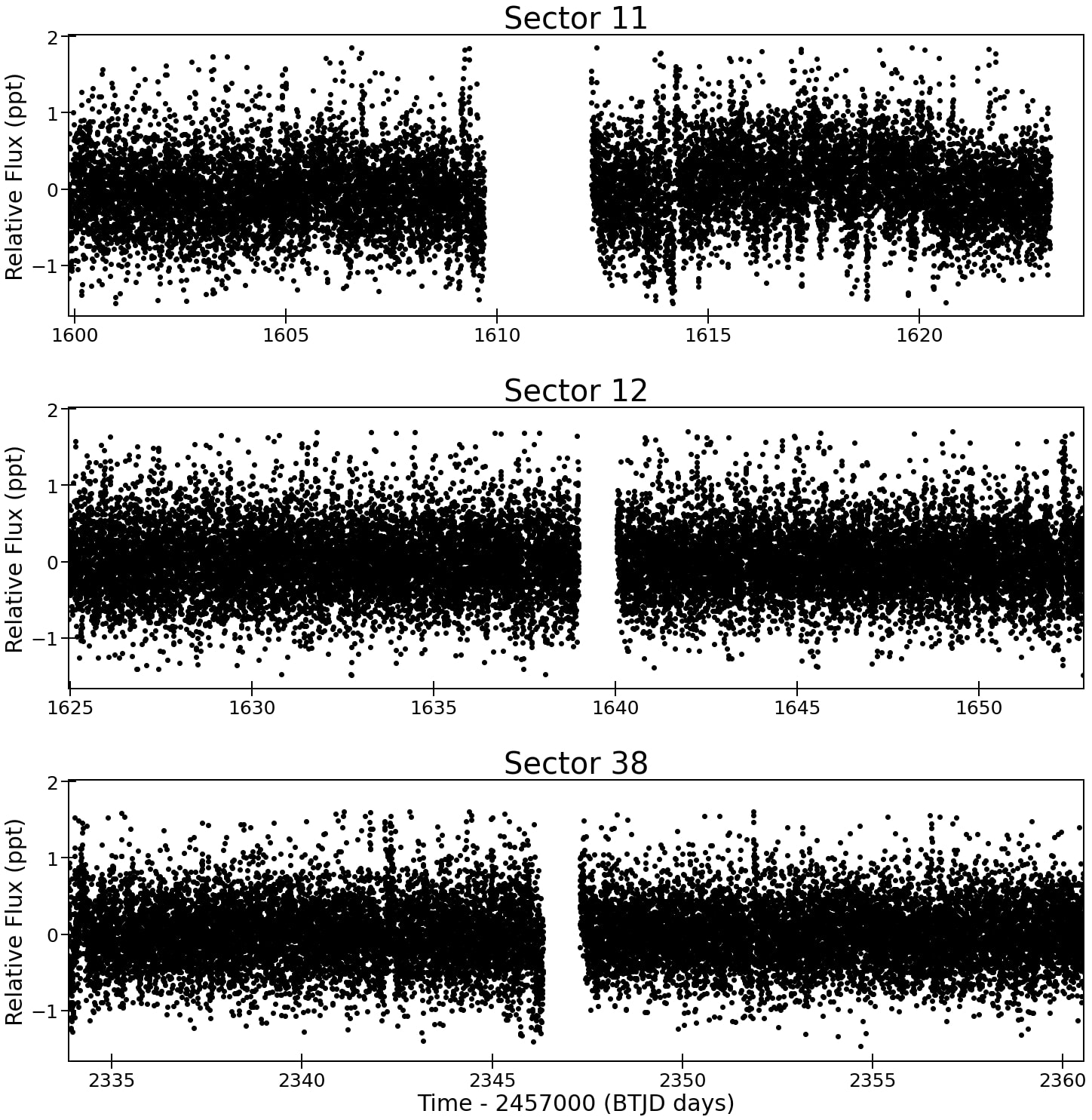}
    \caption{After flare modeling and removal, we are left with these cleaned light curves on which we conduct our planet search.}
    \label{fig:clean-lightcurves}
\end{figure}

The flare model can be seen in Figure \ref{fig:lightcurves} and the cleaned light curve with flare model subtracted (and sigma-clipped) that we use to search for planets is shown in Figure \ref{fig:clean-lightcurves}. There is still some non-Gaussian noise in the data, but it is low enough to facilitate transiting planet searches for low signal-to-noise signals.

\subsection{Planet Searches}
We used two different planet search algorithms to attempt to detect a planet in the Proxima Centauri data. We first used the Transit Least Squares (TLS) software \citep{tls}, with all three sectors of data, and then again with just Sector 11-12. We used the default search parameters aside from changing the mass and radius of the host star to match that of Proxima Centauri. No planetary signals were detected with orbital periods between 1 and 30 days at a significant above 2-$\sigma$ in each of the search methods. Following the methods of \citet{kruse2019}, we also used the Quasiperiodic Automated Transit Search (QATS) planet search method. QATS allows planets to have a transiting-timing variations and still be recovered, but we again did not find any significant signals.

\subsection{Planet Injection}

\begin{figure}
    \centering
    \includegraphics[width=\textwidth]{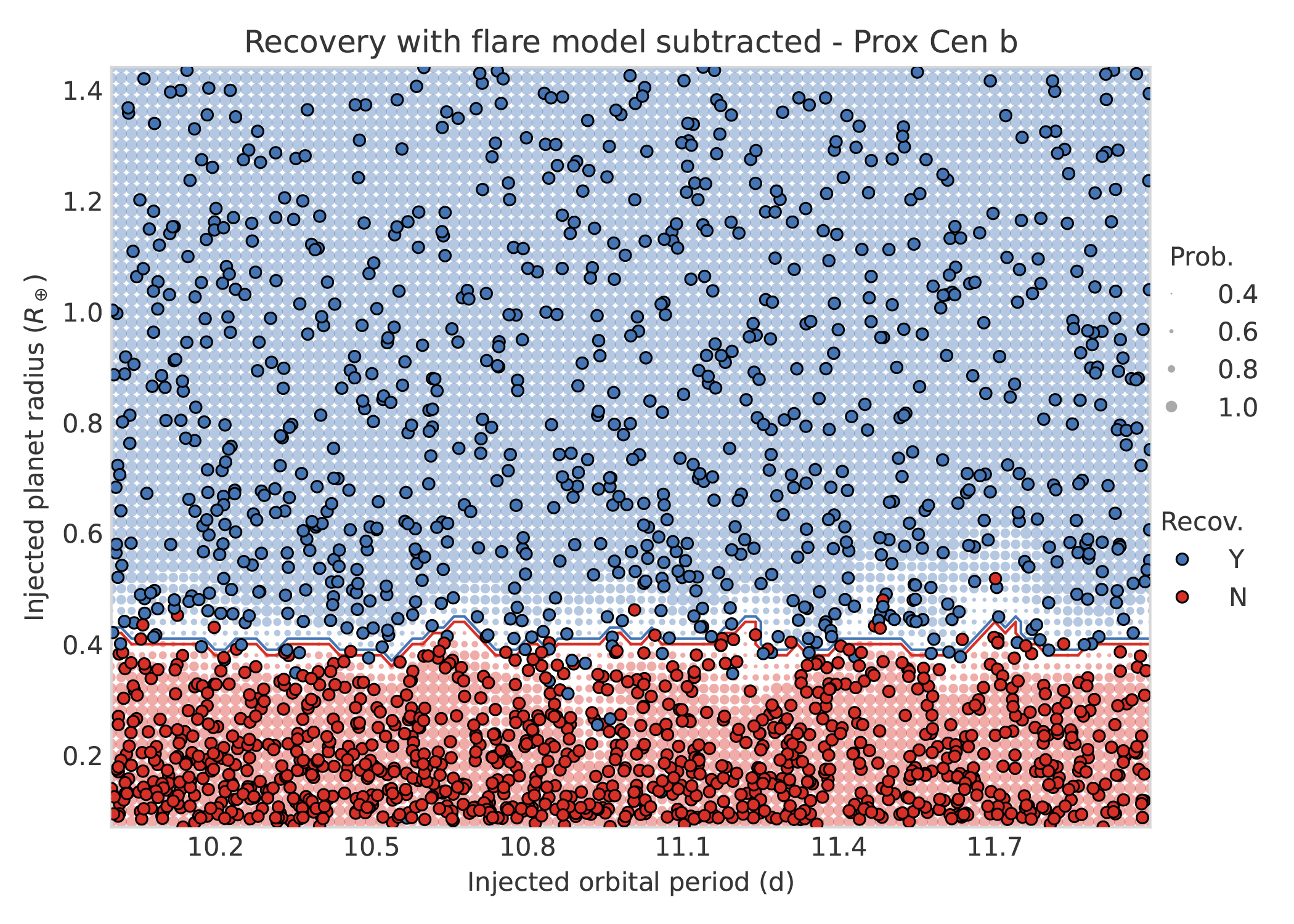}
    \caption{By subtracting a flare model from the light curve of Proxima Centauri, we are able to significantly increase the probability of recovering small planets. We are able to reliably recover planets down to around the radius of Mars across the period range searched, effectively ruling out any transit of Proxima Centauri b.}
    \label{fig:injection-proxb}
\end{figure}

While we did not detect any planets, it is essential that we characterize our sensitivity. To determine the sensitivity of our method, we injected transits into the light curves before any type of flare modeling or detrending. Then, after the transits are injected, we performed the steps described in Section~\ref{sec:flares}. We used \texttt{batman} \citep{batman} to calculate transit models and injected planets that transit the center of the star on circular orbits, with linear orbital ephemerides. The planet size and orbital periods were randomly drawn from log-uniform distributions with radii ranging from 0.07--1.5 R$_\oplus$ and orbital periods from 10--12 days. The transit epoch was randomly drawn from a uniform distribution between zero and the orbital period. The 10--12 day period of the injected planets enables us to test out sensitivity to the finding the 11.2 day orbital period planet Proxima Centauri b. We injected over 3,000 planets and ran the TLS algorithm to search for planets. We used Sectors 11--12 in this search between it speeds up the computation very significantly, and we found that the additional sector only provided limited additional sensitivity to small planets with orbital periods shorter than 25 days. The results of this injection are shown in Figure~\ref{fig:injection-proxb}. We found that we are sensitive to planets larger than approximately 0.4--0.5 R$_\oplus$ across the period-range that that we searched, and every planet larger than 0.55 R$_\oplus$ that we injected was detected. To better quantify our detection probability we used the k-nearest neighbors algorithm implemented in as the \texttt{KNeighborsClassifier} in \texttt{scikit-learn} \citep{scikit-learn}, with k=12, to calculate a decision boundary between the parameter space where planet planets were recovered from where we have no sensitivity. The decision boundary between recovered and not-recovered signals is relatively flat, and has a mean value of 0.40 R$_\oplus$, as shown in Figure~\ref{fig:injection-proxb}.

\section{Discussion}\label{sec:discussion}

Herein, we discuss the implications of flare modeling when it comes to planet detection. In Subsection \ref{sec:disc-flares}, we compare our injection and recovery results with and without modeling the flares. In Subsection \ref{sec:disc-no-transit}, we confirm that we do not detect any evidence of transits of Proxima Centauri b in the three available sectors of TESS data. Finally, in Subsection \ref{disc-HZ}, we look at the rest of the Proxima Centauri Habitable Zone and comment on the possibility of transiting planets in this regime.

\subsection{The consequence of not modeling the flares}\label{sec:disc-flares}
We have shown that by modeling the flares in the Proxima Centauri light curve, we can have sensitivity to very small planets (as small as Mars-sized). However, we have not yet demonstrated that we are more sensitive to small planets because we are modeling the flares. To determine whether there is any improvement we performed the same steps as before, but this time did not attempt to model the flares. We examined two different methods, firstly we simply interactively sigma-clipped out data-points, and secondly we tried using the flare model but clipping out the data points that occur during times that flare occur rather than subtracting the model.

\begin{figure}
    \centering
    \includegraphics[width=\textwidth]{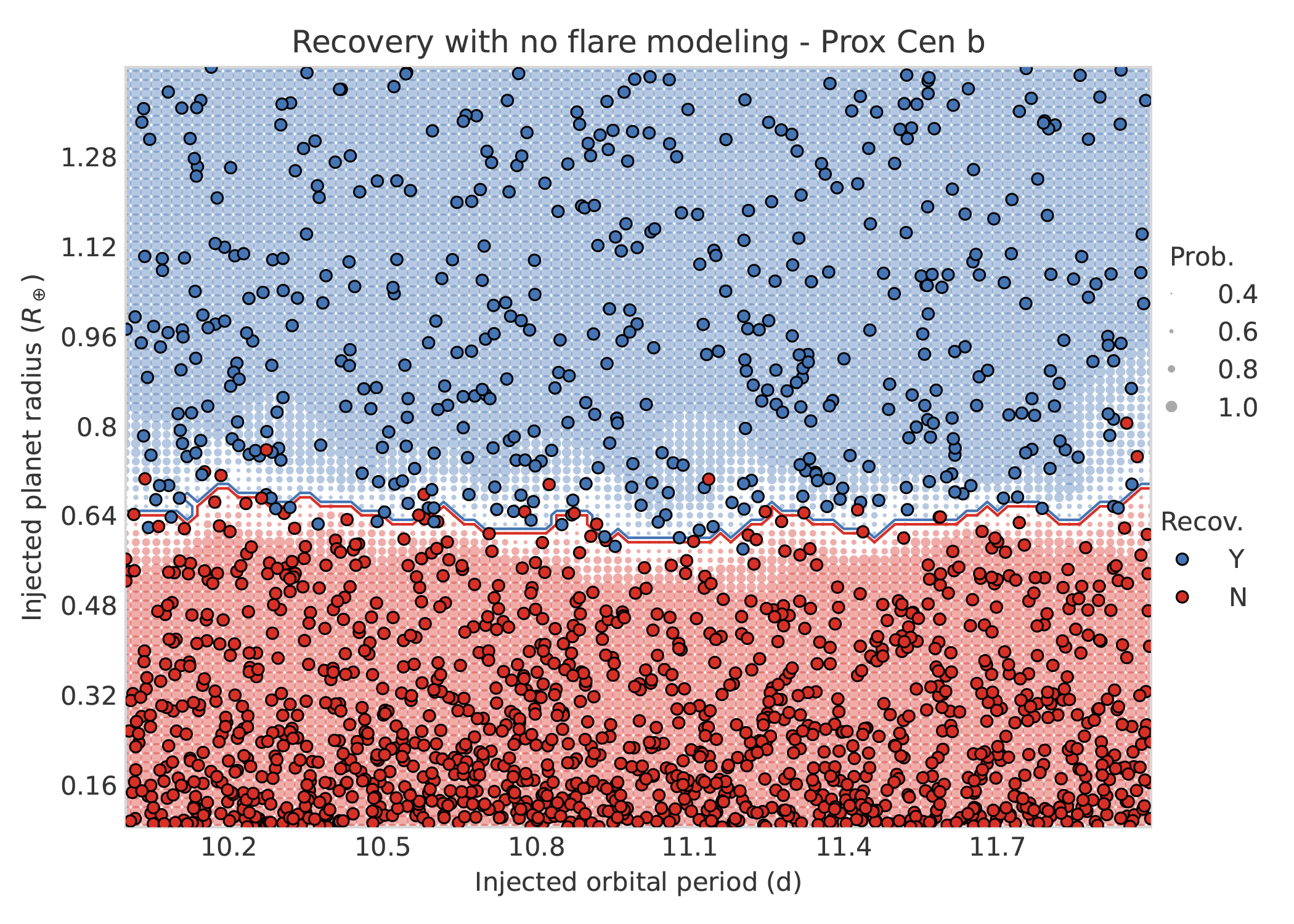}
    \caption{Without accounting for flares in our planet injection and recovery tests, we are less likely to recover the smallest planets. This figure shows the sensitivity to transiting planet when using sigma-clipping of flares, rather than flare model subtraction. Compared to the flare modeling sensitivity is 68\% poorer.}
    \label{fig:injection-proxb-dirty}
\end{figure}

We found that both methods for more naively dealing with the flares resulted in significantly poorer sensitivity to planets. Figure~\ref{fig:injection-proxb-dirty} shows the planets recovered when sigma-clipping the data. The mean of the recovery boundary is 0.67 R$_\oplus$, which is 68\% higher than when we created a model of the flares. The second method of masking data points that are during flares resulted in even much poorer results with regions of period-space with no sensitivity, and should not be used.

\subsection{Proxima Centauri b does not transit}\label{sec:disc-no-transit}
We find no evidence for Proxima Centauri b in TESS data. This is not surprising because previous efforts using different telescopes have been similarly fruitless. 

Using the known minimum mass of Proxima Centauri b (Msini = 1.27 M$_\oplus$), we used the relationship from \citet{chenandkipping} to derive an expected planet radius to be R = 1.08 $\pm$ .14 R$_\oplus$. A 100\% Iron planet would have an expected radius of 0.88 R$_\oplus$ \citep{Zeng2019}. Therefore, given our injection and recovery tests show that no planets larger than 0.4 R$_\oplus$ transit Proxima Centauri at periods between 10--12 days, we are confident that we would recover the signal from Proxima Centauri b if it were to transit.

\subsection{Searching for additional transiting planets in the Proxima Centauri Habitable Zone}\label{disc-HZ}
\begin{figure}
    \centering
    \includegraphics[width=0.48\textwidth]{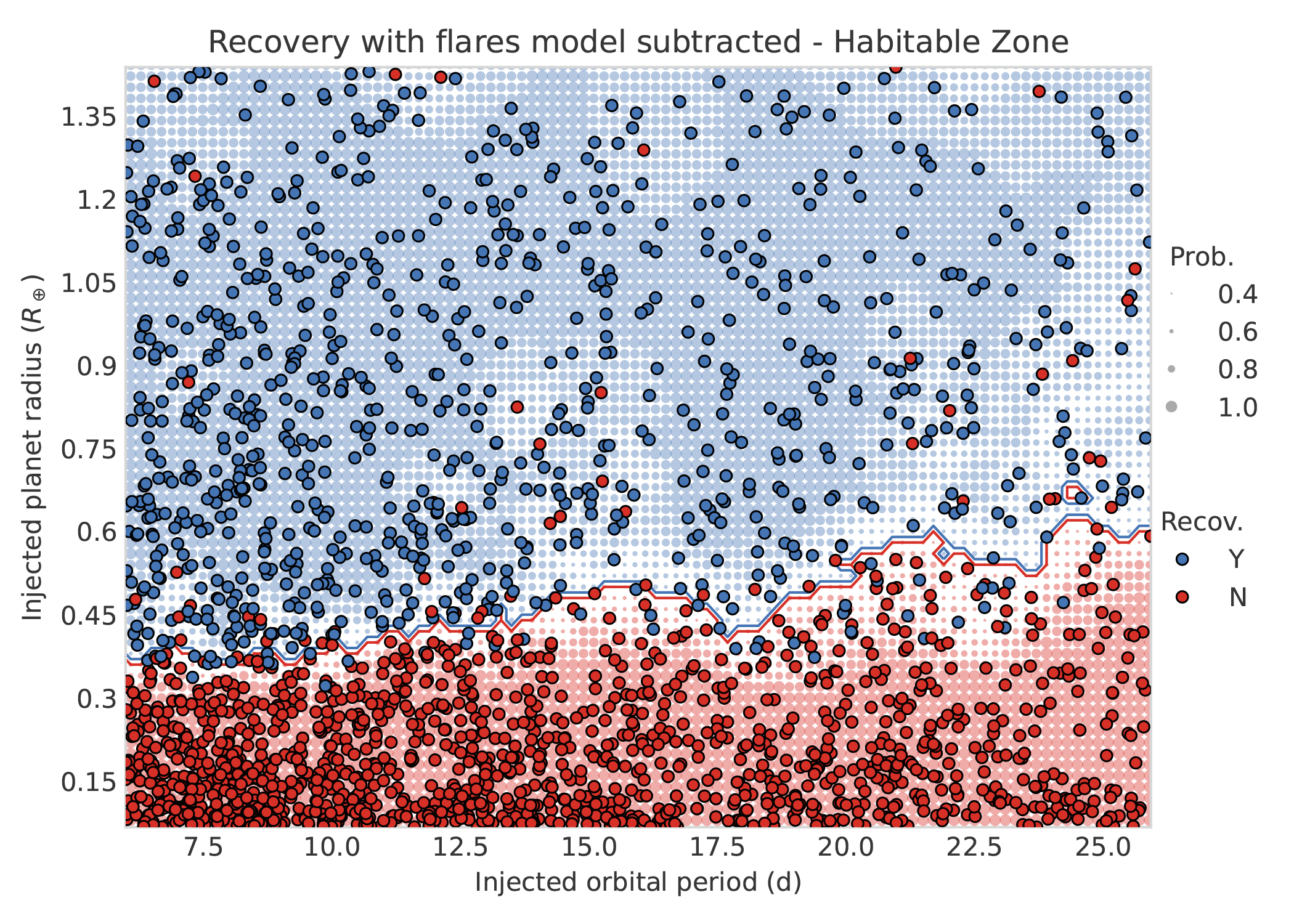} \includegraphics[width=0.48\textwidth]{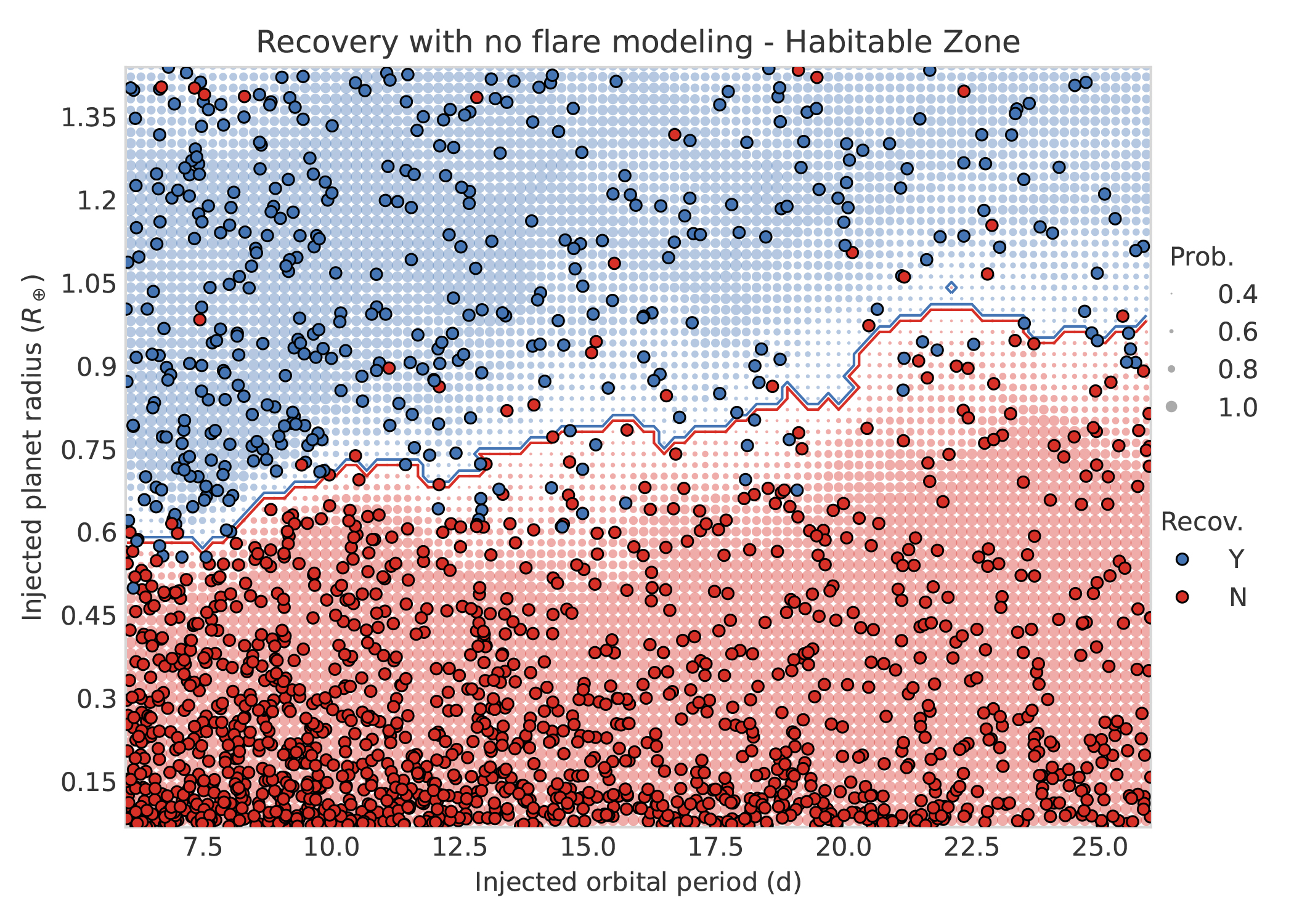}
    \caption{An expanded planet injection and recovery test spanning the entire Habitable Zone of Proxima Centauri. On the left, we show results with the flare model subtracted from the light curve, which provides significant improvement as compared to results without subtracting off the flares, as shown on the right.}
    \label{fig:injection-proxhz}
\end{figure}

We performed our initial search for transiting planets over orbital periods ranging from 1--30 days and did not identify any planets. Using the method demonstrated in the previous sections, we can also quantify additional the sensitivity we have to previous undetected transiting planets in the Proxima Centauri data from TESS. We limited our search to the optimistic Habitable Zone \citep{kopparapu2013} which spans from orbital periods of 6--26.6 days and injected planets with the same sizes, and ran the same recovery simulation. The results of this search are shown in Figure~\ref{fig:injection-proxhz} for the two cases where flares are modeled. We are sensitive to planets as small as 0.35 R$_\oplus$ at the short end of the period range, and 0.6 R$_\oplus$ at long period end. When the flares are just sigma-clipped the sensitivity ranges from 0.6--1.0 R$_\oplus$, demonstrating a 60\% improvement in our sensitivity to small planets over standard methods.

No transiting planets larger than 0.6 R$_\oplus$ orbit within Proxima Centauri's Habitable Zone.

\section{Conclusions}\label{sec:conclusions}

Proxima Centauri is the closest star to the Sun, and one of the best studied stars in the sky. The detection of a planet orbiting our nearest stellar neighbor was a ground-breaking moment for exoplanet science. While unlikely, if Proxima Centauri b were to transit, it would enable new insights into exo-atmospheres through transmission spectroscopy methods. Unfortunately we have confirmed prior results that Proxima Centauri b is not a transiting planet. We rule out any planet larger than 0.4 R$_\oplus$ with an orbital period comparable to Proxima Centauri b.

Modeling the flares resulted in a very significant improvement in the sensitivity to transiting planets (0.4 versus 0.6 R$_\oplus$). By careful treatment of the flares observed in TESS photometry of Proxima Centauri, we were able to place rigorous limits on the probability of transiting planets orbiting the host star.

We also searched for transiting planets in the Habitable Zone of Proxima Centauri, but we again came up short. We showed that no planet smaller than Mars transits the star from our point of view within the entire Habitable Zone.

The methods presented here for detecting transiting planets by modeling the flares were highly effective. Proxima Centauri is an unsually active star, and therefore our method is especially successful for this target. However, it is likely to be generally applicable for all active stars. With the ongoing TESS mission as well as planned missions like Plato providing long baseline, high-precision observations, this technique may be extremely valuable for detecting small planets orbiting active host stars. There are many low-mass active nearby stars, and the methods we have presented here are could significantly improve our sensitivity to small planets transiting these stars.

\section*{Conflict of Interest Statement}

The authors declare that the research was conducted in the absence of any commercial or financial relationships that could be construed as a potential conflict of interest.

\section*{Author Contributions}

E.G. conducted the flare detection and wrote the manuscript. T.B. ran the planet injection and recovery tests. E.K. searched for additional planets with QATS. E.Q. and L.W. provided additional guidance, ideas. All authors contributed to the manuscript.

\section*{Funding}
The material is based upon work supported by NASA under award number 80GSFC21M0002. This work was also supported by NASA award 80NSSC19K0315. This work was supported by the GSFC Sellers Exoplanet Environments Collaboration (SEEC), which is funded by the NASA Planetary Science Division’s Internal Scientist Funding Model.

\section*{Acknowledgments}
This paper includes data collected by the TESS mission, which are publicly available from the Mikulski Archive for Space Telescopes (MAST). Funding for the TESS mission is provided by NASA's Science Mission directorate. E.A.G. thanks the LSSTC Data Science Fellowship Program, which is funded by LSSTC, NSF Cybertraining Grant \#1829740, the Brinson Foundation, and the Moore Foundation; her participation in the program has benefited this work. 

This work made use of the following software:

astropy \citep{exoplanet:astropy13,exoplanet:astropy18}, 
IPython \citep{ipython}, 
Jupyter \citep{jupyer}, 
Lightkurve \citep{lightkurve}, 

Matplotlib \citep{matplotlib},

NumPy \citep{numpy}, 

PyMC3 \citep{exoplanet:pymc3}, 
scikit-learn \citep{scikit-learn},

Theano \citep{exoplanet:theano}, 
xoflares \citep{xoflares}
.

\section*{Data Availability Statement}
The datasets analysed for this study can be found in the Mikulski Archive for Space Telescopes (MAST) \url{https://archive.stsci.edu/}.

\bibliographystyle{frontiersinSCNS_ENG_HUMS} 
\bibliography{refs}

\end{document}